\documentclass[sn-aps]{sn-jnl}

\jyear{2022}%

\theoremstyle{thmstyleone}%
\theoremstyle{thmstyletwo}%

\theoremstyle{thmstylethree}%

\usepackage{changepage}
\usepackage{multirow}
\usepackage{amssymb}
\usepackage{array}
\usepackage{epsfig}
\usepackage{color}
\usepackage{graphicx}
\usepackage{subfig}
\usepackage{tabularx}
\usepackage{hyperref}
\usepackage{caption}
\usepackage{float}
\usepackage{siunitx}
\usepackage{threeparttable}

\usepackage{enumerate}
\usepackage{amsmath}
\usepackage{url}

\begin{document}

\title[PDE-Constrained Registration]{PDE-constrained shape registration to characterize biological growth and morphogenesis from imaging data}


\author*[1]{\fnm{Aishwarya} \sur{Pawar}}\email{pawarar@purdue.edu}

\author[2]{\fnm{Linlin} \sur{Li}}\email{li2212@purdue.edu}

\author[3]{\fnm{Arun K.} \sur{Gosain}}\email{ArGosain@luriechildrens.org}

\author[2]{\fnm{David M.} \sur{Umulis}}\email{dumulis@purdue.edu}

\author*[1,2]{\fnm{Adrian} \sur{Buganza Tepole}}\email{abuganza@purdue.edu}

\affil*[1]{\orgdiv{School of Mechanical Engineering}, \orgname{Purdue University}, \orgaddress{\street{585 Purdue Mall}, \city{West Lafayette}, \postcode{47907}, \state{Indiana}, \country{USA}}}

\affil[2]{\orgdiv{Weldon School of Biomedical Engineering}, \orgname{Purdue University}, \orgaddress{\street{206 S Martin Jischke Dr}, \city{West Lafayette}, \postcode{47907}, \state{Indiana}, \country{USA}}}

\affil[3]{\orgdiv{Lurie Children's Hospital}, \orgname{Northwestern University}, \orgaddress{\street{225 East Chicago Ave}, \city{Chicago}, \postcode{60611}, \state{Illinois}, \country{USA}}}


\abstract{We propose a PDE-constrained shape registration algorithm that captures the deformation and growth of biological tissue from imaging data. Shape registration is the process of evaluating optimum alignment between pairs of geometries through a spatial transformation function. 
We start from our previously reported work, which uses 3D tensor product B-spline basis functions to interpolate 3D space. Here, the movement of the B-spline control points, composed with an implicit function describing the shape of the tissue, yields the total deformation gradient field. The deformation gradient is then split into growth and elastic contributions. The growth tensor captures addition of mass, i.e. growth, and evolves according to a constitutive equation which is usually a function of the elastic deformation. Stress is generated in the material due to the elastic component of the deformation alone. The result of the registration is obtained by minimizing a total energy functional which includes: a distance measure reflecting similarity between the shapes, and the total elastic energy accounting for the growth of the tissue. We apply the proposed shape registration framework to study zebrafish embryo epiboly process and tissue expansion during skin reconstruction surgery. We anticipate that our PDE-constrained shape registration method will improve our understanding of biological and medical problems in which tissues undergo extreme deformations over time.}

\keywords{Adaptive Refinement, Surface Registration, Tissue Expansion, Truncated Hierarchical B-splines, Zebrafish Epiboly, Growth and Remodeling, Isogeometric Analysis}



\maketitle

\section{Introduction}
\label{sec1}
Shape registration is the process by which two or more geometries are aligned and deformed to achieve accurate correspondence. Given a pair of geometries, namely the source and target geometry, it is desirable to construct spatial transformations which are regular, smooth and result in one-to-one maps between shapes. These smooth and invertible spatial transformations are also known as diffeomorphisms \cite{ashburner2007fast,beg2005computing}. Registration based on free-form deformation (FFD) using B-splines has emerged recently as a powerful tool in image analysis due to the smoothness and local control of B-spline basis functions \cite{sederberg1986free,szeliski1997spline, rueckert1999nonrigid,tustison2009directly, tustison2013explicit}. In our earlier work \cite{pawar2019joint}, we developed a registration framework based on B-splines, which allowed smooth, diffeomorphic and large deformations of 3D space. One of our main contributions in that earlier work was local refinement using truncated hierarchical B-splines (THB-splines) to maximize computational efficiency. THB-splines were used to automatically refine regions where significant deformation was expected. In addition to the spatial transformations being diffeomorphic, it is desirable to simultaneously achieve physically realistic maps  ~\cite{veress2005deformable, mang2018pde}. Many physical processes involve the solution of partial differential equations (PDEs). Therefore, a key need in image analysis is the integration of PDE constraints on segmentation and registration frameworks.

Imaging data for biomedical applications include confocal microscopy, magnetic resonance images (MRI), and computer tomography (CT) scans, among others. In some situations, longitudinal data is available, i.e. images of the same tissue at multiple time points. For example, longitudinal MRI scans of cancerous tumors or aneurysms are used clinically to determine the timing of intervention \cite{meier2016clinical,chien2019unruptured}. Longitudinal imaging data is also used to understand fundamental processes of biological systems, such as quantification of embryo morphogenesis from confocal microscopy images \cite{keller2008reconstruction}. Registration of longitudinal imaging data enables quantitative analysis of shape changes in development or disease. Imaging data alone, however, is only one piece of the puzzle. A range of physical phenomena accompany the geometric changes seen in 3D images. Chiefly, tissues deform and carry stress in response to applied forces and constraints. Then, mechanical cues lead to tissue adaptation through addition of mass -referred to as growth- and remodeling of material properties \cite{tepole2011growing,taber1995biomechanics}. 

Computational models of growth and remodeling have been developed over the past couple of decades to better characterize tissue biomechanics and mechanobiology \cite{cowin2004tissue, tepole2017computational}. A common approach to model growth and remodeling is through finite element simulations \cite{goktepe2010generic, himpel2005computational}. Unfortunately, uncertainty in material parameters and boundary conditions prevents accurate representation of a biological system, and a simplified model is typically used instead. For instance, in our previous work on skin growth in tissue expansion we have modeled skin as an idealized flat piece of tissue \cite{tepole2011growing}. 

There exists a gap at the intersection of computational modeling of growth and remodeling biomechanics and the registration of longitudinal imaging data of biological systems. To fill this gap, we propose a novel shape registration framework to capture the deformation of biological tissue from imaging data while satisfying the finite growth framework within continuum mechanics \cite{goriely2007definition}. For the registration framework we start from our previously reported work using THB-splines ~\cite{pawar2016adaptive, pawar2018dthb3d, pawar2019joint}. The tissue is considered as a hyperelastic solid and hyperelastic strain energy is the constraint to model physically realistic deformations, e.g. \cite{veress2005deformable}. Different from previous work, in our framework we capture both growth and remodeling of tissues by including the multiplicative split of the deformation gradient field. The split of the deformation gradient into growth and elastic contributions is akin to finite deformation plasticity \cite{lee1969elastic}. Linear momentum balance is sought, with the stress being a function of the elastic deformation only. The growth tensor, on the other hand, obeys an ordinary differential equation (ODE) encoding morphogenesis and mechanobiology information. We apply the novel shape registration framework to study the growth of biological tissues in two applications: skin expansion and zebrafish embryo growth.

\section{Methods}
We consider source and target geometries as $\mathcal{B}_1$ and $\mathcal{B}_2$. Functions $S_1(\textbf{x})$ and $S_2(\textbf{x})$ with $\mathbf{x}\in\mathbb{R}^3$ coordinates of 3D space are implicit representations of the source and target geometries, respectively. Namely, $\mathcal{B}_1 \equiv \{\mathbf{X}\in \mathbb{R}^3 \, \mathrm{s.t.} \, S_1(\mathbf{X})=1\}$ is the source geometry, while $\mathcal{B}_2 \equiv \{\mathbf{Y}\in \mathbb{R}^3 \, \mathrm{s.t.} \, S_2(\mathbf{Y})=1\}$ is the target geometry. The goal of the shape registration framework is to evaluate an optimal spatial transformation mapping $\mathbf{\varphi}(\mathbf{X})$ that is smooth and diffeomorphic, resulting in accurate alignment of the shapes matched: $\mathbf{Y}=\mathbf{\varphi}(\textbf{X})$ which can be checked through the implicit function by achieving $ S_2(\mathbf{\varphi}(\mathbf{X})) = S_1(\mathbf{x}) $.


\begin{figure}[!ht]
\centering
\includegraphics[width=0.9\textwidth]{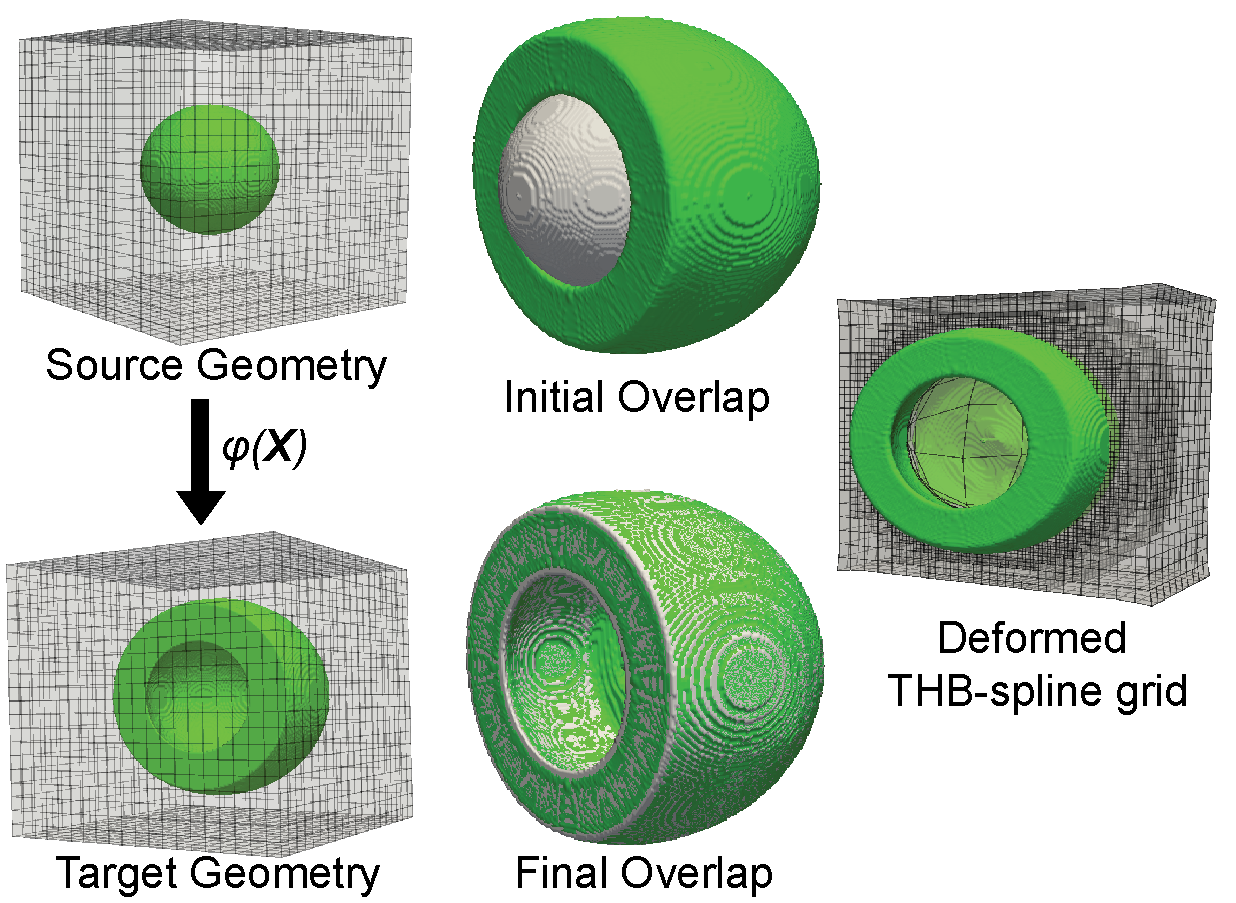}
\caption{Schematic diagram of shape registration. The B-spline grid overlays the source and target geometries. The spatial mapping $\varphi(\mathbf{X})$ is evaluated that corresponds to the best match. Complex and large deformation is captured using THB-spline grid as shown from the initial and final overlap between the geometries.}
\label{fig1:schematic_diagram}
\end{figure} 

\subsection{Spatial Transformation and THB-splines}
Due to the inherent smoothness and local control properties of B-splines, we utilize free-form deformation \cite{sederberg1986free,rueckert1999nonrigid} to evaluate the spatial transformation function. In free-form deformation, the B-spline control grid overlays the 3D space which has coordinates $\textbf{x}$. The grid is deformed through the spatial transformation $\mathbf{\varphi}(\mathbf{x})$. Note that the entire grid deformation is described by $\mathbf{\varphi}(\mathbf{x})$, whereas the points of the source image (the reference body) are only the subset $\mathbf{X}$ that satisfies $S_1(\mathbf{X})=1$.  The initial 3D space is parameterized as 
\begin{equation}
\textbf{x} =\sum_{m=1}^{N_b}\textbf{A}_{m}B_{m,p}(\textbf{u}),
\label{eq_grid}
\end{equation}
where $\textbf{x}$ is the 3D space, $\textbf{A}_{m}$ are the initial set of control points, $B_{m,p}$ are basis functions which are evaluated on a parameter domain $\textbf{u}=[u,v,w]$. In fact, we choose $\mathbf{u}=\mathbf{x}$ as the parameter domain itself, and in that case the control points $\textbf{A}_{m}$ need to be picked as the Greville abscissae \cite{gordon1974b,farin2014curves}. $N_b$ represents the total number of trivariate basis functions. $B_{m,p}(\textbf{u})$ is the tensor product of $p^{th}$ order univariate B-spline basis functions $N_{i,p}(u)$, $N_{j,p}(v)$ and $N_{k,p}(w)$ defined on the open knot vectors $U = \{ u_1, \cdots, u_{n_1+p+1} \}$, $V = \{ v_1, \cdots, v_{n_2+p+1} \}$ and $W = \{ w_1, \cdots, w_{n_3+p+1} \}$ spanning the 3D space in $u$, $v$ and $w$ directions, respectively. Here $n_1$, $n_2$ and $n_3$ are the number of univariate basis functions in each parametric direction. Each B-spline $N_{i,p}(u)$ in this parametric domain has local support defined as $\mathrm{supp}(N_{i,p}(u)) = [u_i, u_{i+p+1}]$. Given that initial representation of 3D space, the  spatial transformation function can be written as
\begin{equation}
\textbf{y} =\mathbf{\varphi}(\mathbf{x})= \sum_{m=1}^{N_b}\textbf{P}_{m}B_{m,p}(\textbf{x}),
\label{eq_spatialmap}
\end{equation} where $\textbf{P}_{m}$ are the new locations of the control points. Note, therefore, that the functions are still defined in terms of the initial parameterization, but the movement of the control points now leads to a different set of coordinates $\mathbf{y}$. The use of the lower case $\mathbf{x}$ and $\mathbf{y}$ is used to encompass the entire deformation of the initial 3D space parameterization, but we emphasize again that the source geometry we are interested are only the points $\mathbf{X}$ such that $S_1(\mathbf{X})=1$ and the target geometry are the points $\mathbf{Y}$ such that $S_2(\mathbf{Y})=1$.

We carry out local refinement using truncated hierarchical B-splines (THB-splines) \cite{giannelli2012thb} in this manuscript for the evaluation of spatial transformation function. Through local refinement, only the regions near the shape boundaries are refined to capture highly localized deformations \cite{pawar2018dthb3d,pawar2016adaptive,xie2004image}. We explain the implementation of local refinement using THB-splines through an example of a univariate B-spline basis function for two refinement levels. Consider univariate B-splines at two parametric domains, a coarser ($\Omega_0$) and finer domain ($\Omega_1$). We can represent each basis function at the coarser level $l$ as the linear combination of B-splines at finer refinement level $l+1$ \cite{bornemann2013subdivision}. In addition to the coarser level B-spline which is substituted by finer B-splines, the basis functions with partial support in the local support of the coarser B-spline are truncated. Such refinability property is given as 

\begin{equation}
    N_{i,p}^{l}(u) = \sum_{j=1, \mathrm{supp}(N_{j,p}^{(l+1)}(u)) \notin \Omega_1 }^{n_c} W_{i,j} N_{j,p}^{(l+1)}(u),
\label{eq_thb}
\end{equation} where $W_{i,j}$ are the subdivision coefficients determined using the knot insertion algorithm and $n_c$ is the number of children B-splines. To determine the refinement criterion in the registration framework, we first evaluate $I_g = \lvert \nabla (S_2^{l} (\mathbf{\varphi}(\mathbf{X}))-S_1(\textbf{X})) \rvert$ at the center point of each B-spline control grid element. For each B-spline, we evaluate the average of $I_{g}$ in its support domain, denoted as $G_{j}$. We refine the B-spline basis function which satisfies $G_{j} > \rho \, G_{\mathrm{mean}}$, with $\rho$ being a parameter that controls the amount of refinement and $G_{\mathrm{mean}}$ being the average value of $I_g$ over the entire domain \cite{pawar2016adaptive}.

\subsection{Registration Framework}
\begin{figure}[!ht]
\centering
\includegraphics[width=\textwidth]{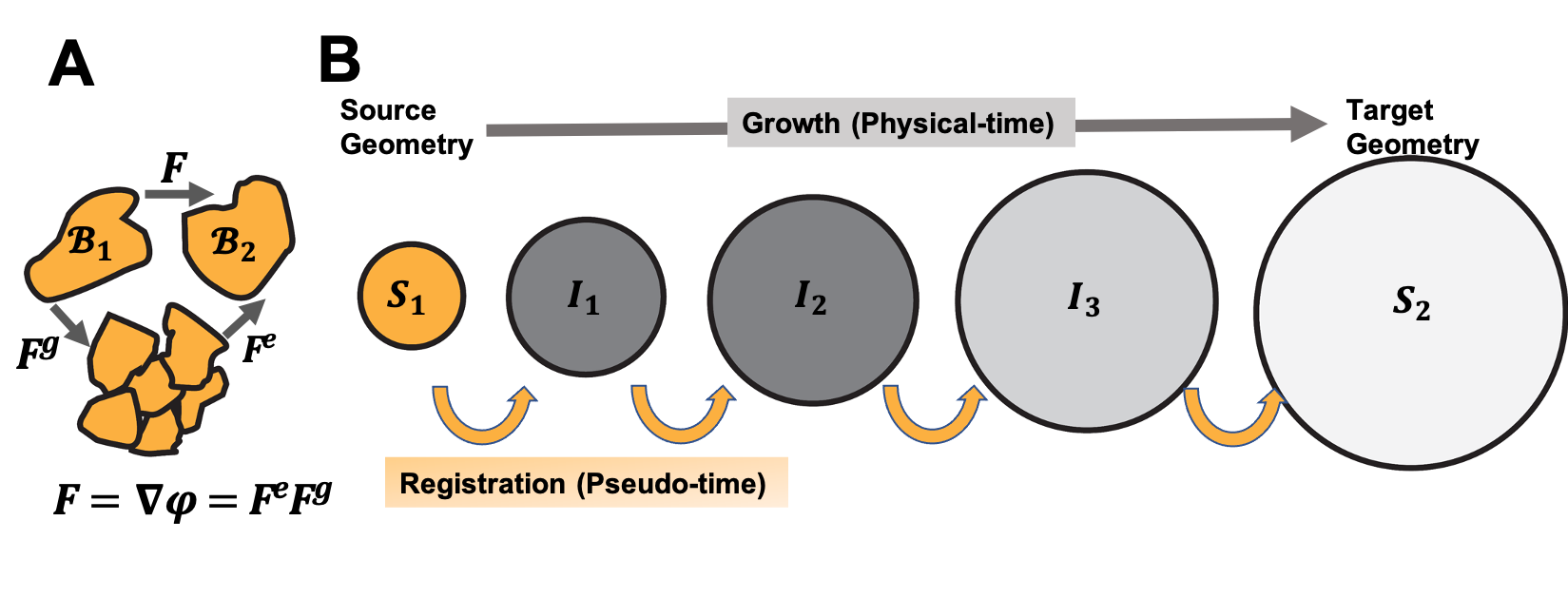}
\caption{A. Schematic of multiplicative split of the deformation gradient $\mathbf{F}$ into growth ($\mathbf{F^{g}}$) and elastic part ($\mathbf{F^e}$). B. Schematic diagram showing the staggered solution of the elastic deformation and growth problems within the shape registration framework. The deformation of the source to the the target is divided into intermediate stages. This can be achieved by either creating linear interpolations ($I_1, I_2, \cdots I_n$) of the geometry from source to target, or increasing the image error penalty $\alpha$ monotonically, e.g. linearly.}
\label{fig:growth_schematic}
\end{figure}

Resuming from the definition of geometry above, $\textbf{X} \in \mathcal{B}_1$ denotes the initial configuration, $\mathbf{\varphi}(\mathbf{X})$ maps the deformation of the initial configuration to the target configuration $\mathcal{B}_2$. The total deformation gradient is defined as $\textbf{F} = \nabla_\mathbf{X} \mathbf{\varphi}(\mathbf{X})$. The multiplicative split into growth and elastic components is carried out in order to model growth \cite{rodriguez1994stress}. This is akin to plasticity \cite{lee1969elastic},
\begin{equation}
    \mathbf{F} = \mathbf{F}^{\mathbf{e}} \mathbf{F}^{\mathbf{g}},
\label{eq_split_deformation}
\end{equation} where $\mathbf{F}^{\textbf{e}}$ is the deformation tensor associated with hyperelastic deformation and $\mathbf{F}^{\textbf{g}}$ is the deformation tensor associated with growth. The determinant of total deformation gradient can also be similarly split as $J = J^{e} J^{g}$, where $J^{e}$ and $J^{g}$ correspond to the elastic and growth volume changes respectively as shown in Fig. \ref{fig:growth_schematic}A. For the examples shown in this manuscript, the biological tissue is considered to be neo-Hookean hyperelastic material, but this can easily be exchanged for other material models used for soft tissue \cite{holzapfel2001biomechanics}. The strain energy of the neo-Hookean solid is defined as  


\begin{equation}
    \Psi^{e} = \frac{1}{2} \mu \, (\mathrm{tr}(\textbf{C}^\textbf{e})-3) - \mu \ln{(J^e)} + \frac{\lambda}{2} (\ln{(J^e)})^2,
\label{eq_hyperelastic}
\end{equation} where $\mu$ and $\lambda$ are Lame's parameters. $\textbf{C}^{e}$ is the elastic right Cauchy-Green tensor defined as $\textbf{C}^\textbf{e} = \textbf{F}^\textbf{e T} \textbf{F}^\textbf{e}$.

In the proposed method, the weak form of mechanical equilibrium is evaluated using the B-spline basis functions. Our approach follows existing finite element implementations of growing tissue \cite{tepole2011growing}, but within the isogeometric analysis framework \cite{pawar2016adaptive}. Given $\mathbf{\varphi}(\mathbf{X})$, the elastic energy functional ($E_{REG}$) is defined as 
\begin{equation}
E_{REG} = \int_{\Omega}  S_1(\mathbf{x}) \Psi^{e}(\mathbf{F}^e) \, \mathrm{d}\Omega,
\label{eq_ereg}
\end{equation} where $\Omega$ is the entire space considered in terms of the coordinates $\mathbf{x}$. The use of $S_1(\mathbf{x})$ inside of the integral essentially computes the elastic strain energy as if it was integrated over $\mathcal{B}_1$ rather than the whole space $\Omega$. Note that the hyperelastic strain energy is a function of the elastic part of the deformation. However, the elastic deformation $\mathbf{F}^e$ is a function of both the total deformation transformation $\mathbf{\varphi}(\mathbf{X})$, as well as the growth deformation $\mathbf{F}^g$. Therefore, to evaluate the elastic deformation we need to also specify how growth changes over time. Usually, growth is prescribed as a local change through an ordinary differential equation (ODE) that describes the rate of growth \cite{cowin2004tissue,goktepe2010generic}. In particular, assuming volumetric growth, the growth tensor can be expressed in terms of a single scalar \cite{himpel2005computational}
\begin{equation}
    \mathbf{F}^g = \theta^g \mathbf{I},
\label{eq_thetag_1}
\end{equation} where $\mathbf{I}$ is the identity and the total growth is now in terms of the scalar $\theta^g$. The ODE for growth change over time is 
\begin{equation}
    \dot{\theta}^g = k(\theta^e-\theta^{\mathrm{crit}})\, ,
\label{eq_thetag_2}
\end{equation} where $k$ is a rate parameter and $\theta^{\mathrm{crit}}$ is a parameter that specifies that growth only takes place beyond a critical value of deformation. Note that the growth rate is actually coupled to the elastic deformation, in this case to the elastic volume change, $\theta^e = J^e = \mathrm{det}\mathbf{F}^e$. Thus, the discretization and integration algorithms will determine how to solve the registration problem. In this work, the energy in Eqn. (\ref{eq_ereg}), is evaluated at fixed growth. In other words, given a fixed field for $\theta^g(\mathbf{X})$, the elastic deformation can be considered a function of the total deformation alone through $\mathbf{F}^e = \mathbf{F}\mathbf{F}^{g-1}$, with $\mathbf{F}=\nabla_\mathbf{X}\mathbf{\varphi}(\mathbf{X})$ showing the explicit dependence on the deformation. Separately, given a total deformation as fixed, i.e. for a fixed deformation $\mathbf{\varphi}(\mathbf{X})$, the growth field can be updated point-wise with a forward Euler scheme
\begin{equation}
    \theta^g_{n+1} (\mathbf{x})= \theta^g_{n} (\mathbf{x}) + \bigtriangleup t \, S_1(\textbf{x}) k \left[J(\mathbf{x}){\theta^g_{n}(\mathbf{x})}-1\right],
\end{equation} where the subscript $n$ denotes the previous time step, time $t$, and $n+1$ is the time step $t+\bigtriangleup t$. Thus, the solution of the registration problem including elastic deformation and growth is done in a staggered manner as indicated in Fig. \ref{fig:growth_schematic}B. Given the elastic energy and growth problems, the registration problem can be introduced by considering the energy coming from the mismatch between the deformed source and target geometries 

\begin{equation}
    E(\mathbf{\varphi}(\textbf{x})) =   \int_\Omega \! (S_1(\textbf{x}) - S_2(\mathbf{\varphi}(\mathbf{x})))^2 \, \mathrm{d}\Omega \, .
\label{eq_image}
\end{equation}

This is the standard image energy in our previous work and other registration work \cite{pawar2016adaptive,pawar2018dthb3d}. Usually, regularization terms are added to the image energy mismatch. These regularization terms penalize changes in the first and second derivatives of the transformation map, which are reminiscent of an elastic energy penalty \cite{leng2013medical}. We also employ a regularization over the entire deformation $\mathbf{\varphi}(\mathbf{x})$, as will be seen soon. Considering the image mismatch energy, the elastic energy introduced in Eqn. (\ref{eq_ereg}), and the regularization of the complement of $\mathcal{B}_1$, the total energy functional reads
\begin{equation}
E(\varphi(\textbf{x})) =   \alpha \int_\Omega \! (S_1(\textbf{x}) - S_2(\mathbf{\varphi}(\mathbf{x})))^2 \, \mathrm{d}\Omega +\beta \int_\Omega  S_1(\textbf{x}) \Psi^{e} \, \mathrm{d}\Omega  + \beta_1 \int_\Omega  (1-S_1(\textbf{x})) \Psi^{e} \, \mathrm{d}\Omega.
\label{eq_total}
\end{equation} 
$\alpha$ is the weighting parameter associated with the image mismatch error, $\beta$ and $\beta_1$ are the weighting parameters for the hyperelastic energy constraint and regularization over the rest of the space. Both integrals are over $\Omega$ but the multiplication of the integrands by either $S_1(\textbf{x})$ or $1-S_1(\textbf{x})$ effectively leads to integrals over $\mathcal{B}_1$ and its complement in $\Omega$.

The minimization of the energy functional is carried out using $L{2}$ gradient flow algorithm \cite{leng2013medical}. The control points are updated using a dynamic scheme in the direction of the variation of the energy functional to  changes in $\mathbf{\varphi}(\mathbf{x})$ \cite{jia2015novel} 

\begin{equation}
\begin{aligned}
&\frac{\partial \textbf{P}_i}{\partial \tau} = {} \alpha \int_\Omega \! 2 (S_1(\textbf{x}) - S_2(\mathbf{\varphi}))\nabla S_2(\mathbf{\varphi})B_i(\textbf{x}) \, \mathrm{d}\Omega
\\& + 2 \, \beta \,  \int_\Omega \! S_1(\textbf{x}) \mathbf{S}:\mathbf{\delta}\mathbf{E_i} \, \mathrm{d}\Omega + 2 \, \beta_1 \,  \int_\Omega \! (1-S_1(\textbf{x})) \mathbf{S}:\mathbf{\delta}\mathbf{E_i} \, \mathrm{d}\Omega,
\end{aligned}
\label{eq_dynamic}
\end{equation} where $\mathbf{P}_i$ is a particular control point and $\tau$ denotes a pseudo-time variable for the dynamic relaxation of the energy functional. On the right-hand side, $B_i(\mathbf{x})$ is the corresponding basis function, $\mathbf{S}$ is the second Piola Kirchhoff stress field, and $\delta \mathbf{E}_i$ is the variation of the Euler Lagrange strain tensor for the corresponding variation of the $i^{th}$ control point. We remark again that the minimization of Eqn. (\ref{eq_total}) is done at fixed growth, and the growth ODE is evaluated in a staggered manner with respect to the registration steps, as shown in Fig. \ref{fig:growth_schematic}. 

\section{Benchmark Examples}
\label{sec3}
In this section, we demonstrate the results of the registration framework on benchmark geometries. Here, we validate the results of our registration framework with analytical solutions before moving on to realistic applications. 
\subsection{Benchmark Examples Without Growth}
\begin{figure}[!htb]
\centering
\includegraphics[width=\textwidth]{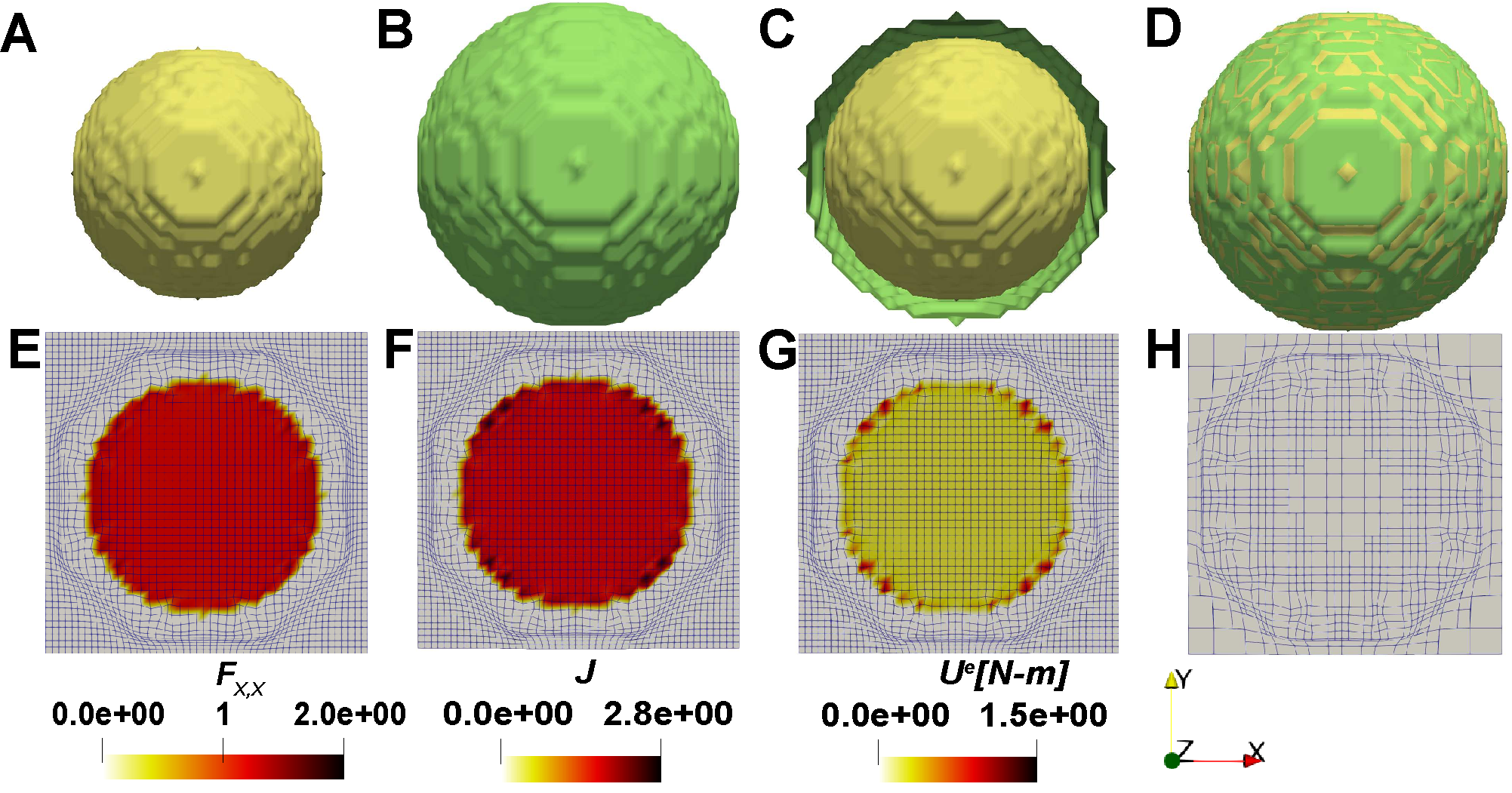}
\caption{Benchmark example of uniform volumetric sphere expansion with hyperelastic regularization without growth. A: sphere image (source) of radius $\SI{15}{\meter}$, B: sphere image (target) of radius $\SI{18}{\meter}$, C: initial difference between the source and target images, D: final difference between the source and target images, E: the deformation gradient along X-direction ($F_{x,x}$) contour, F: local volumetric change ($J$) contour, G: strain energy contour ($U^{e}$) and H: THB-spline grids after third refinement level.}
\label{fig:benchmark_hyperelastic_sphere}
\end{figure}

\begin{figure}[!htb]
\centering
\includegraphics[width=\textwidth]{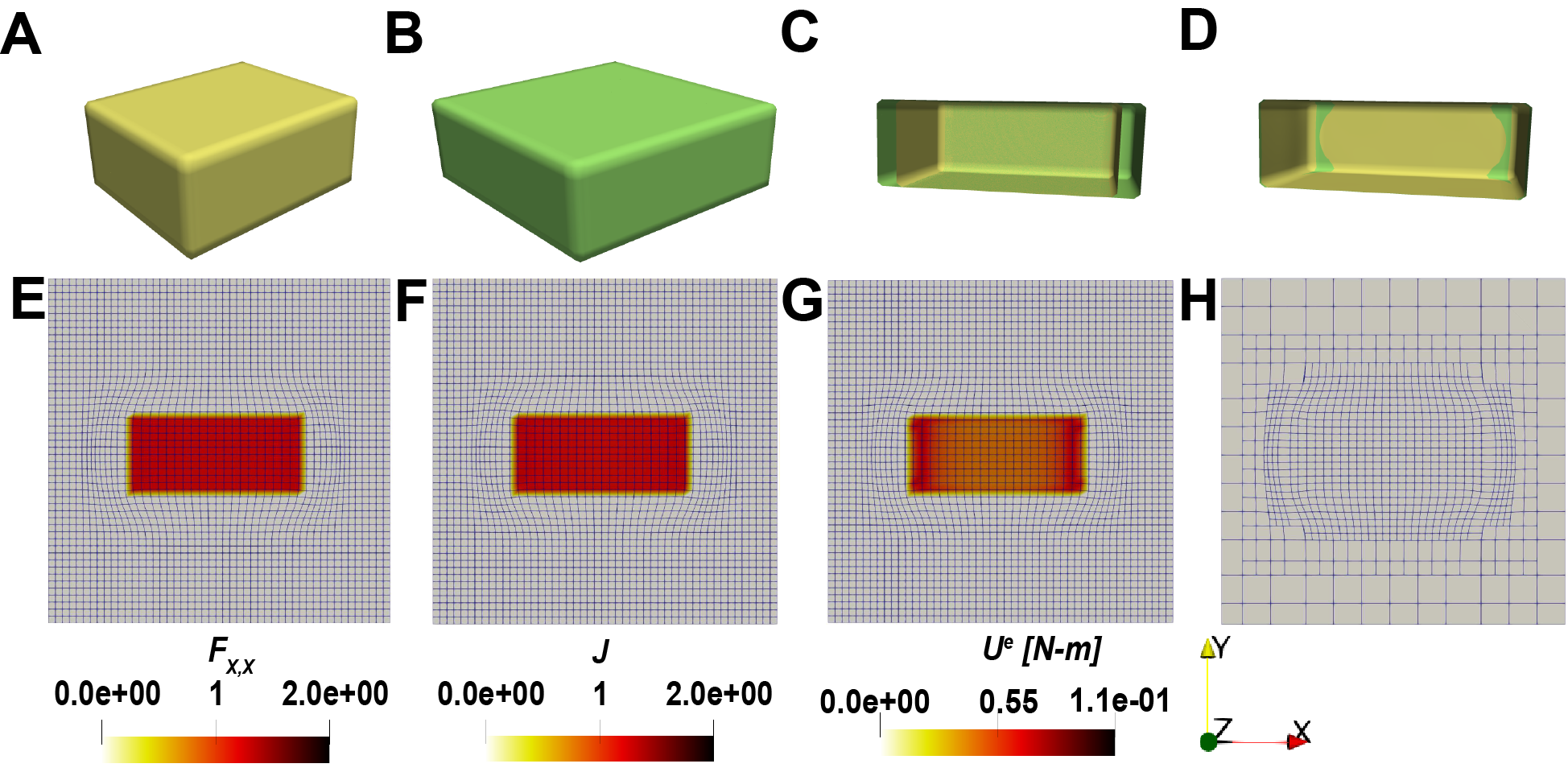}
\caption{Benchmark example of uniform uniaxial expansion of a rectangular plate with hyperelastic regularization without growth. A: plate image (source) of dimension $19\times19\times9$ m, B: plate image (target) of dimension $23\times19\times9$ m, C: initial difference between the source and target images, D: final difference between the source and target images, E: the deformation gradient along X-direction ($F_{x,x}$) contour, F: local volumetric change ($J$) contour, G: strain energy contour ($U^{e}$) and H: THB-spline grids at the end of registration.}
\label{fig:benchmark_hyperelastic_uniaxial}
\end{figure}

\begin{figure}[!htb]
\centering
\includegraphics[width=\textwidth]{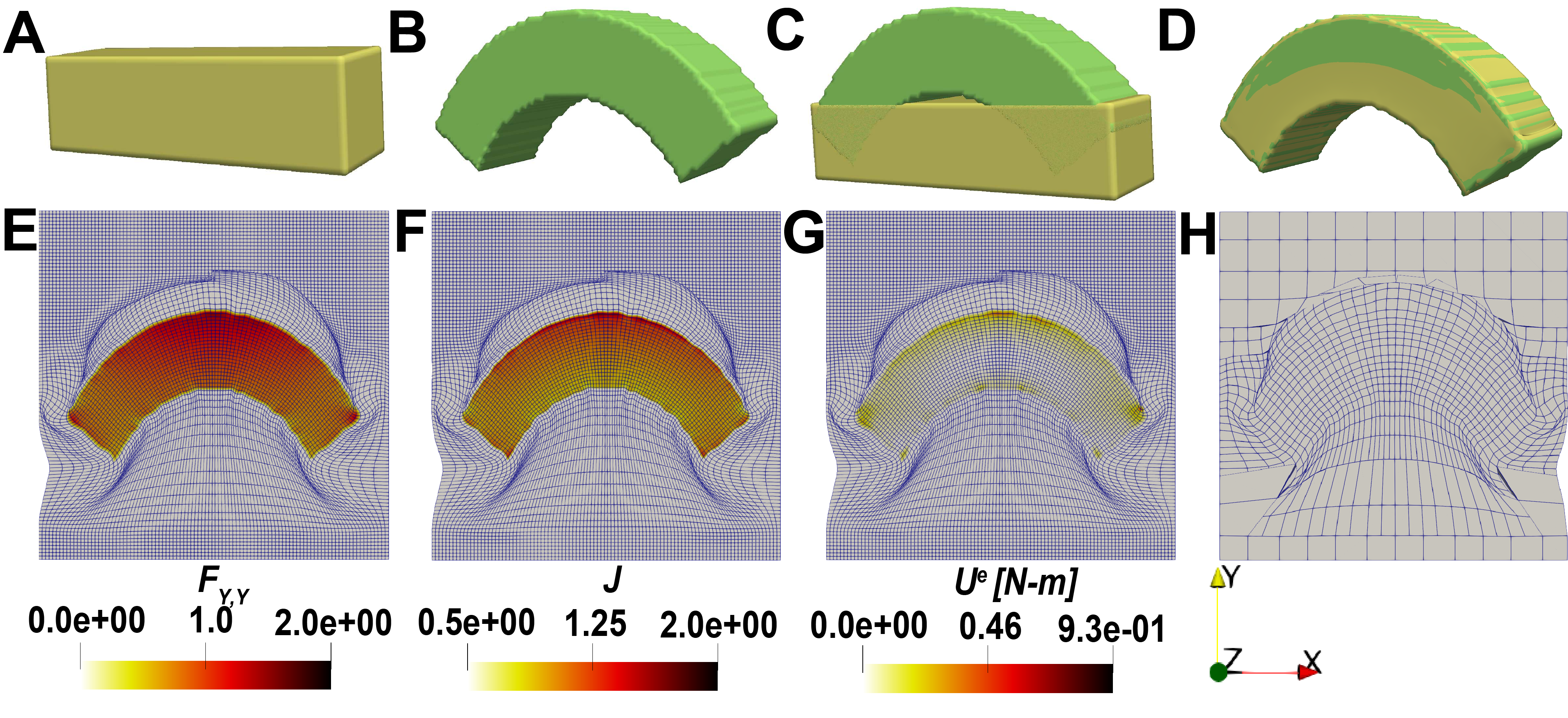}
\caption{Benchmark example of bending deformation of a rectangular plate with hyperelastic regularization without  growth. A: plate image (source), B: plate image (target) after bending deformation is applied, C: initial difference between the source and target images, D: final difference between the source and target images, E: the deformation gradient along Y-direction ($F_{y,y})$ contour, F: local volumetric change ($J$) contour, G: strain energy contour ($U^{e}$) and H: THB-spline grids at the end of registration.}
\label{fig:benchmark_hyperelastic_bending}
\end{figure}

In Figs. \ref{fig:benchmark_hyperelastic_sphere}-\ref{fig:benchmark_hyperelastic_bending}, we first carry out shape registration by considering only hyperelastic deformation without growth. Following cases are shown: Fig. \ref{fig:benchmark_hyperelastic_sphere} homogeneous volumetric expansion of the sphere,  Fig. \ref{fig:benchmark_hyperelastic_uniaxial} uniaxial expansion of a rectangular plate, and Fig. \ref{fig:benchmark_hyperelastic_bending} bending of a rectangular plate. The implicit function representing the 3D solid for Figs. \ref{fig:benchmark_hyperelastic_sphere}-\ref{fig:benchmark_hyperelastic_uniaxial} is evaluated on a grid resolution of $50\times50\times50$ pixels and the initial B-spline control grid has $10\times10\times10$ elements. The grid resolution for the implicit function in Fig. \ref{fig:benchmark_hyperelastic_bending} is $100\times100\times100$ pixels and the initial B-spline control grid has $12\times12\times12$ elements. The Lame parameters $\mu$ and $\lambda$ are both equal to $\SI{1}{\pascal}$. Local  refinement is carried out on three refinement levels. We set a higher value of $\rho$ for increasing refinement levels to prevent introducing large number of control points at higher refinement levels. 

In Fig. \ref{fig:benchmark_hyperelastic_sphere}A, the radius of source geometry is $\SI{15}{\meter}$ while the target geometry has a radius of $\SI{18}{\meter}$ shown in Fig. \ref{fig:benchmark_hyperelastic_sphere}B. $\alpha$ is set as $8$ intially and doubled every refinement level. $\beta$ and $\beta_1$ are set as $1$ and $0$, respectively. The pseudo-time step for dynamic relaxation $\Delta \tau$ is set as $0.005$. The analytical solution of the volumetric deformation of the sphere has deformation gradient components ($F_{x,x}$, $F_{y,y}$ and $F_{z,z}$) equal to $1.2$. The local volumetric change is uniform and equal to $1.728$. The initial overlap between source and target images is depicted in Fig. \ref{fig:benchmark_hyperelastic_sphere}C, while the result of registration is shown in Fig. \ref{fig:benchmark_hyperelastic_sphere}D. From the contours of deformation gradient component $F_{x,x}$ and local volumetric change $J$ shown in Fig. \ref{fig:benchmark_hyperelastic_sphere}E,F, we can see that we achieve homogeneous deformation within most of the region inside the sphere and the values of $F_{x,x}$ and $J$ at the center are equal to $1.188$ and $1.677$, respectively, as desired. The contour of strain energy is depicted in Fig. \ref{fig:benchmark_hyperelastic_sphere}G, further showing the mostly uniform deformation. Lastly, the deformed grid after the third refinement level is shown in  Fig. \ref{fig:benchmark_hyperelastic_sphere}H, where it can be seen that refinement of the mesh is necessary at the boundaries of the sphere.

The uniaxial test case is summarized in Fig. \ref{fig:benchmark_hyperelastic_uniaxial}. The initial plate image has dimensions $19\times19\times9$ m, while the target image of the plate is $23\times19\times9$ m, seen in Fig. \ref{fig:benchmark_hyperelastic_uniaxial}A,B. The initial and final state of the registration algorithm is depicted in Fig. \ref{fig:benchmark_hyperelastic_uniaxial}C,D where it can be seen that the source image has been successfully deformed to match the target. A uniform deformation is expected in the interior of the domain, which can be seen in the contours for one of the deformation gradient components and the volume change, Fig. \ref{fig:benchmark_hyperelastic_uniaxial}E,F. The strain energy contour in Fig. \ref{fig:benchmark_hyperelastic_uniaxial}G shows some small regional variation with slightly more strain energy at the boundaries.


In Fig. \ref{fig:benchmark_hyperelastic_bending}A, the source geometry is a rectangular plate. For the target geometry we calculate the new positions ($\textbf{v}_1$) by applying the deformation at each vertex in the source geometry ($\textbf{v}=[x,y,z]$) as $\textbf{v}_1 = [a\,y\,\sin(a\,x)\,+\,\frac{1}{a}\,y\,\sin(a\,x), a\,y\,\cos(a\,x)\,+\,\frac{1}{a}\,y\,\cos(a\,x), z ]$, and we set $a = 1$ to generate the image in Fig. \ref{fig:benchmark_hyperelastic_bending}B. We set the initial value of $\alpha$ as $1$ and double it every $500$ iterations. Since we have large deformation in this benchmark example, we set $\beta$ to a higher value of $2$. A systematic test of the regularization parameters is covered later on in the manuscript. The final overlap between the images after registration is shown in Fig. \ref{fig:benchmark_hyperelastic_bending}D. Based on the deformation applied on the plate, the bottom layer undergoes compression while the top layer undergoes tension. This can be seen in the deformation gradient component $F_{y,y}$ in Fig. \ref{fig:benchmark_hyperelastic_bending}E. There is a radial increase in the local volumetric change from the bottom to the top layer. At a particular radial distance, the local volumetric change remains uniform, and this matches well with a pure bending deformation. There are some boundary effects that are noticeable in the strain energy contour in Fig. \ref{fig:benchmark_hyperelastic_bending}G where it can be seen that there is excessive distortion of the mesh right at the sharp corners of the two images. Thus, registration of other geometries with sharp corners undergoing significant displacement might require special treatment in the future. 

\subsection{Parameter Tuning}
The parameter $\beta$ controls the relative weight of the hyperelastic energy in the solid with respect to the energy of the image mismatch. Increasing this parameter can result in slow convergence and higher registration error, while a small value can lead to unrealistic movement of the control points during registration and overlapping of control grids, resulting in high strain energy that is slow to converge. Intuitively, when the variation of the energy is considered, Eqn. (\ref{eq_dynamic}), the energy from the image mismatch leads to an applied external pressure or traction at the boundary of the elastic body. The variation of the hyperelastic energy leads to the weak form of linear momentum balance in Eqn. (\ref{eq_dynamic}), i.e. internal forces due to the deformation of the elastic body. If $\mathcal{B}_1$ is too stiff relative to the applied force from the image mismatch, which is effectively controlled by $\beta$, the tractions obtained from the image mismatch are unable to deform the body. The opposite case, when $\mathcal{B}_1$ is too soft relative to the image mismatch residual, leads to excessively large tractions at the boundary  $\partial \mathcal{B}_1$ which compromise the stability and convergence of the scheme. 

\begin{figure}[!ht]
\centering
\includegraphics[width=\textwidth]{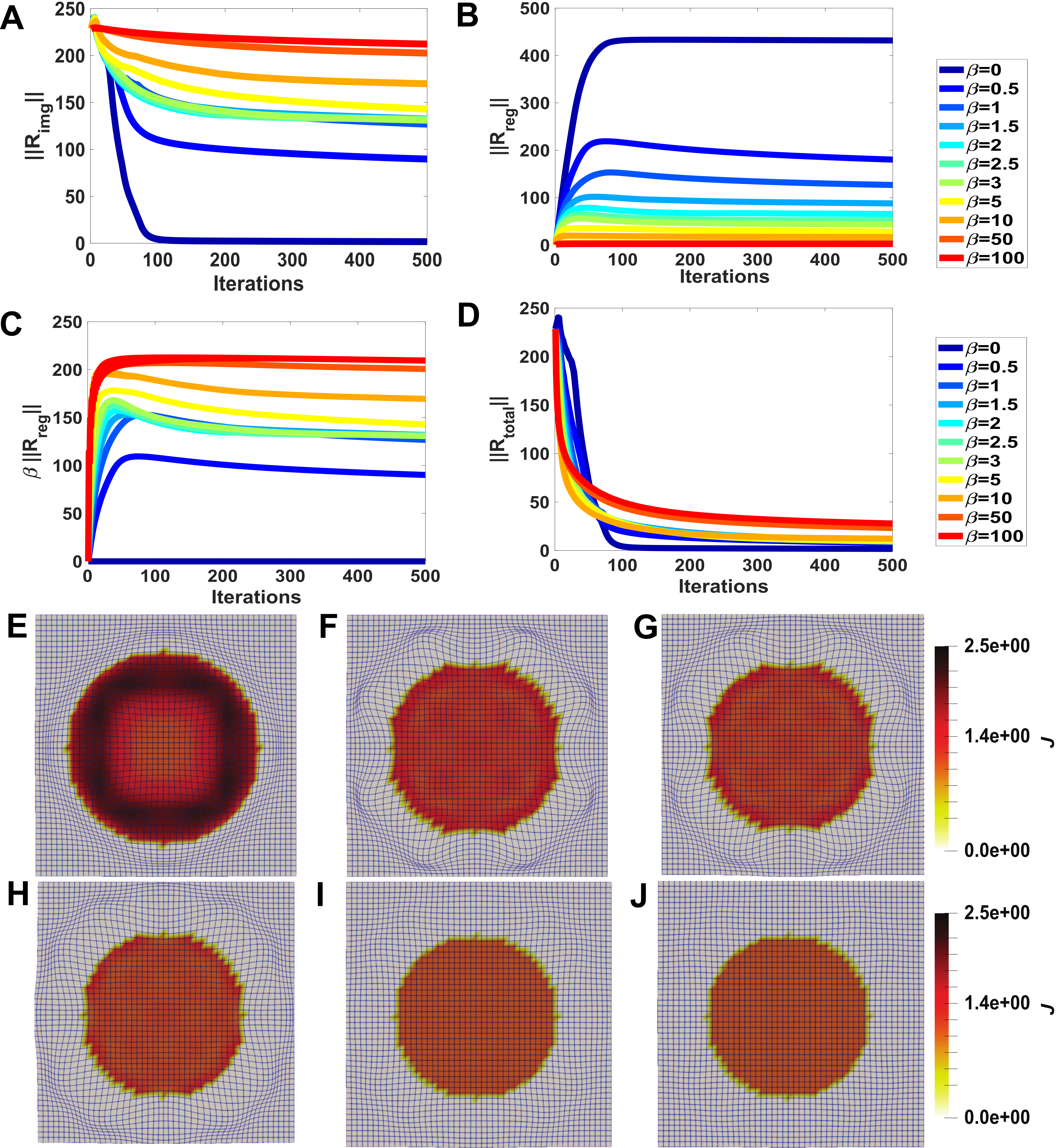}
\caption{Residual plots for the volumetric expansion of the sphere as initially shown in Fig. \ref{fig:benchmark_hyperelastic_sphere} but exploring the change in the hyperelastic regularization parameter $\beta$ ranging from $0$ to $100$. A: the norm of the residual due to image error ($ \vert \vert R_{img} \vert \vert$), B: the norm of the residual due to hyperelastic strain energy ($ \vert \vert R_{reg} \vert \vert$), C: the norm of the residual due to hyperelastic strain energy multiplied with the parameter $\beta$ ($\beta \, \vert \vert R_{reg} \vert \vert$), D: the norm of the residual of the total energy ($\vert \vert R_{total} \vert \vert$). E-K: Local volumetric change contour plots for $\beta$ equal to $0$, $1.5$, $2.5$, $3$, $50$ and $100$.}
\label{fig:residuals_beta}
\end{figure} 

In Fig. \ref{fig:residuals_beta}, the residuals associated with the image mismatch energy, strain energy, and the total energy for different values of $\beta$ ranging from 0 to 100 are plotted. The contours associated with local volumetric change are shown in Fig. \ref{fig:residuals_beta}. As can be seen in the Fig. \ref{fig:residuals_beta}A, we show the residuals associated with the image mismatch energy as we increase $\beta$. For $\beta = 0$ the image residual drops very quickly as there is absolutely no constraint on the type of deformation that is admissible. However, we can see from the local volumetric change contour in Fig. \ref{fig:residuals_beta}E that only the outer region is deformed. Correspondingly, the hyperelastic strain energy residual for $\beta=0$ reaches its maximum value in Fig. \ref{fig:residuals_beta}B because this deformation is not the expected uniform deformation from momentum balance. As we increase the value of $\beta$, the registration leads to a uniform deformation inside the sphere, as desired. However, as explained, there is a trade-off. For a very high $\beta$ value shown in Fig. \ref{fig:residuals_beta}J, there is essentially no deformation. In principle, any $\beta>0$ should lead to the uniform deformation that is expected in this benchmark case, as long as the registration is perfectly achieved. However, if $\beta>0$ but small, the dynamic update can be slow to converge. The  optimum $\beta$ should be small enough to allow for deformation of the solid and minimization of the image energy. Concurrently, $\beta$ should be high enough to prevent initial excessive distortion and allow for rapid stabilization to a steady state. For the sphere case, and the shear modulus $\mu$ chosen, an optimal range of $\beta$ is $2-3$. In this range, $\beta \in [2,3]$, the residuals in Fig. \ref{fig:residuals_beta}E converge to the same value within 500 iterations and the $J$ contour is uniform inside the sphere see Fig. \ref{fig:residuals_beta}G. 



\subsection{Benchmark Example With Growth}
\begin{figure}[!htb]
\centering
\includegraphics[width=\textwidth]{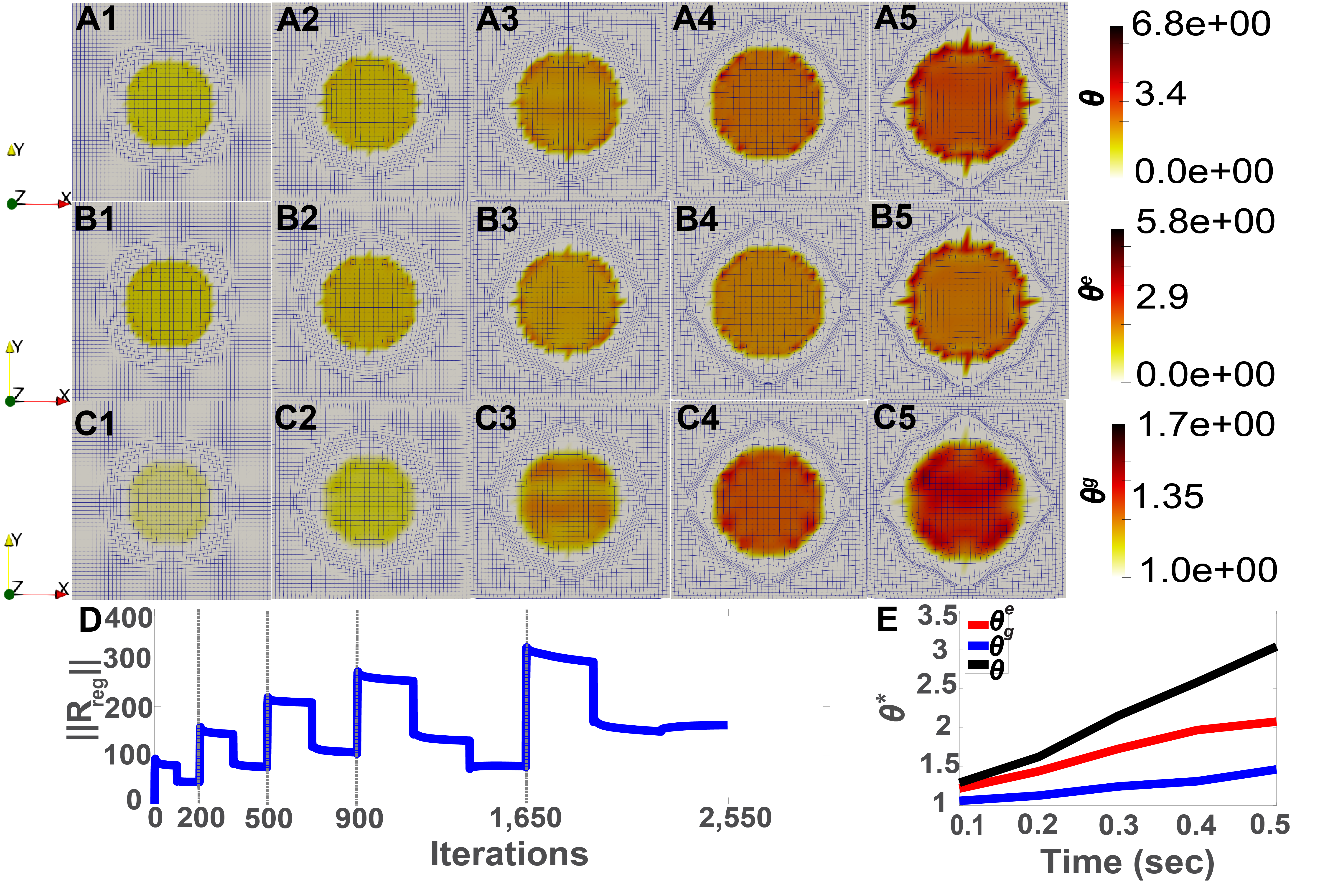}
\caption{Homogeneous volumetric expansion of sphere from radius $10$ m to radius $15$ m combining both hyperelastic and growth deformation. The local volumetric change contours associated with total ($\theta$), elastic ($\theta^{e}$) and growth deformation ($\theta^{g}$) for each linearly interpolated stage are shown in A1-A5, B1-B5 and C1-C5, respectively. The residual curve for hyperelastic strain energy associated with homogeneous volumetric expansion of sphere example is shown in D and the plot of total deformation ($\theta$), elastic deformation ($\theta^{e}$), and growth deformation ($\theta^{g}$) at the center of the sphere with respect to time is shown in E.}
\label{fig:benchmark_growth}
\end{figure}

In Fig. \ref{fig:benchmark_growth}, we carry out shape registration of the uniform sphere expansion, similar to Fig. \ref{fig:benchmark_hyperelastic_sphere}, but also considering volumetric growth. We show the homogeneous volumetric expansion of a sphere with the addition of volumetric growth from radius of $10$ and $15$ m. Because growth needs to be integrated over time, we split the total deformation into multiple stages;   the radius is increased by 1 m in each stage. Growth is observed over a time duration of 5 seconds. The growth rate $k$ is set as $\SI{2}{\sec\tothe{-1}}$ and time step for growth update is set as $\bigtriangleup t = 0.1$ s. $\beta$ and $\beta_1$ are set as $1$ and $2.5$, respectively. Here we are adding higher regularization in the region outside the evolving sphere so that there is not much overlap of grids and its accumulation over interpolations. We remark that the source geometry in each stage remains fixed while the target is linearly interpolated.    

The total volume change ($\theta$) value at the center of the sphere increases from 1 to 3.04 as the sphere radius increases. For each stage, the field is fairly uniform inside the sphere, with some small artifacts at the boundary, see Fig. \ref{fig:benchmark_growth}A1 to A5. The elastic part of the total volume change ($\theta^{e}$) is shown in Fig. \ref{fig:benchmark_growth}, from B1 to B5, while the growth field ($\theta^{g}$) is depicted in Fig. \ref{fig:benchmark_growth} from C1 to C5. Note how the total deformation increases directly based on the registration, i.e. the total change in the sphere radius is directly driven by the increasing size of the target. However, at the end of each interpolation we update the growth field by point-wise integration of the growth equation. Growth, thus, increases over time, but lags behind the total deformation. The elastic deformation field in Fig. \ref{fig:benchmark_growth}B1-B5 is such that $ \theta =\theta^e \theta^g$. 

The residuals associated with the hyperelastic strain energy at all the interpolated stages ($ \vert \vert R_{reg} \vert \vert $) are shown in Fig. \ref{fig:benchmark_growth}D. Note how the residuals evolve discontinuously over the iterations for two reasons. First, for a given target, the parameter $\alpha$ is adjusted at multiple stages to gradually drive the image registration. Second, once registration is converged for a given target, the growth field is updated and the target is then also updated before resuming the iterations. In the end, we are only interested in the deformation field and growth fields at the end of the registration for each interpolation of the target. This is shown in Fig. \ref{fig:benchmark_growth}E, where the total deformation, elastic deformation, and growth deformation are plotted over time for a point at the center of the sphere. As pointed out in the contours, the overall volume change follows what is expected from the gradual change in the target from 10 to 15 $m$. The growth $\theta^g$ lags with respect to the total deformation, as it needs to satisfy the ODE in Eqn. (\ref{eq_thetag_2}). Lastly, the elastic deformation is such that $\theta =\theta^e \theta^g$ is satisfied.

\section{Modeling Epiboly in Zebrafish Embryos}
\label{sec4}
Embryonic development involves cell proliferation and resulting tissue growth which results in large deformations over time. The initial major morphogenetic movement during the gastrulation stage of embryonic development in some organisms is termed epiboly where the blastoderm grows and covers the yolk. In zebrafish, epiboly involves an organized movement of embryonic cells between 4.3 and 10 hours post fertilization (hpf) during which a layer of epithelial cells, are also referred to as the Enveloping Layer (EVL), spreads and covers the yolk cell \cite{bruce2016zebrafish, lepage2010zebrafish, campinho2013tension}. 
At the start of the epiboly process, the blastoderm, a single multilayer of cells is located at the animal pole of the embryo on top of the yolk cell. As proliferation takes place, the EVL thins and increases in area \cite{campinho2013tension}. The yolk syncytial layer (YSL) stays in contact with the yolk cell, causing spherical spreading observed in microscopy images such as in \cite{lepage2010zebrafish,bruce2016zebrafish}. In addition to cell proliferation, other mechanisms that contribute to the epiboly process are the polymerization of actin filaments at the leading edge of the EVL, and myosin-driven contraction on the actin belt that forms at junction between the EVL and the YSL \cite{campinho2013tension,behrndt2012forces}. 

We are interested in capturing the continuous change in shape during zebrafish embryo development, particularly during epiboly, with our shape registration framework. The dataset consists of light sheet microscopy images of early stage zebrafish embryo development from \cite{vladimirov2014light, wu2021automatic}. Additionally, we analyzed the cell proliferation data from \cite{keller2008reconstruction} to develop an accurate and physically-realistic spatial mapping between pairs of datasets of cell positions captured through in \textit{vivo} imaging at different stages of the epiboly process. The positions of cell nuclei ranging from 100 to 1,450 minutes post fertilization (mpf) at 90 sec intervals were collected from whole-mount live light-sheet microscopy images \cite{keller2008reconstruction}. By imposing that the deformation across the different stages of epiboly has to satisfy the momentum balance of an elastic body as well as the growth given by cell division, we seek to learn the spatial distribution of elastic deformation that is expected during this crucial stage in embryo development. 

\begin{figure}[!ht]
\centering
\includegraphics[width=1.01\textwidth]{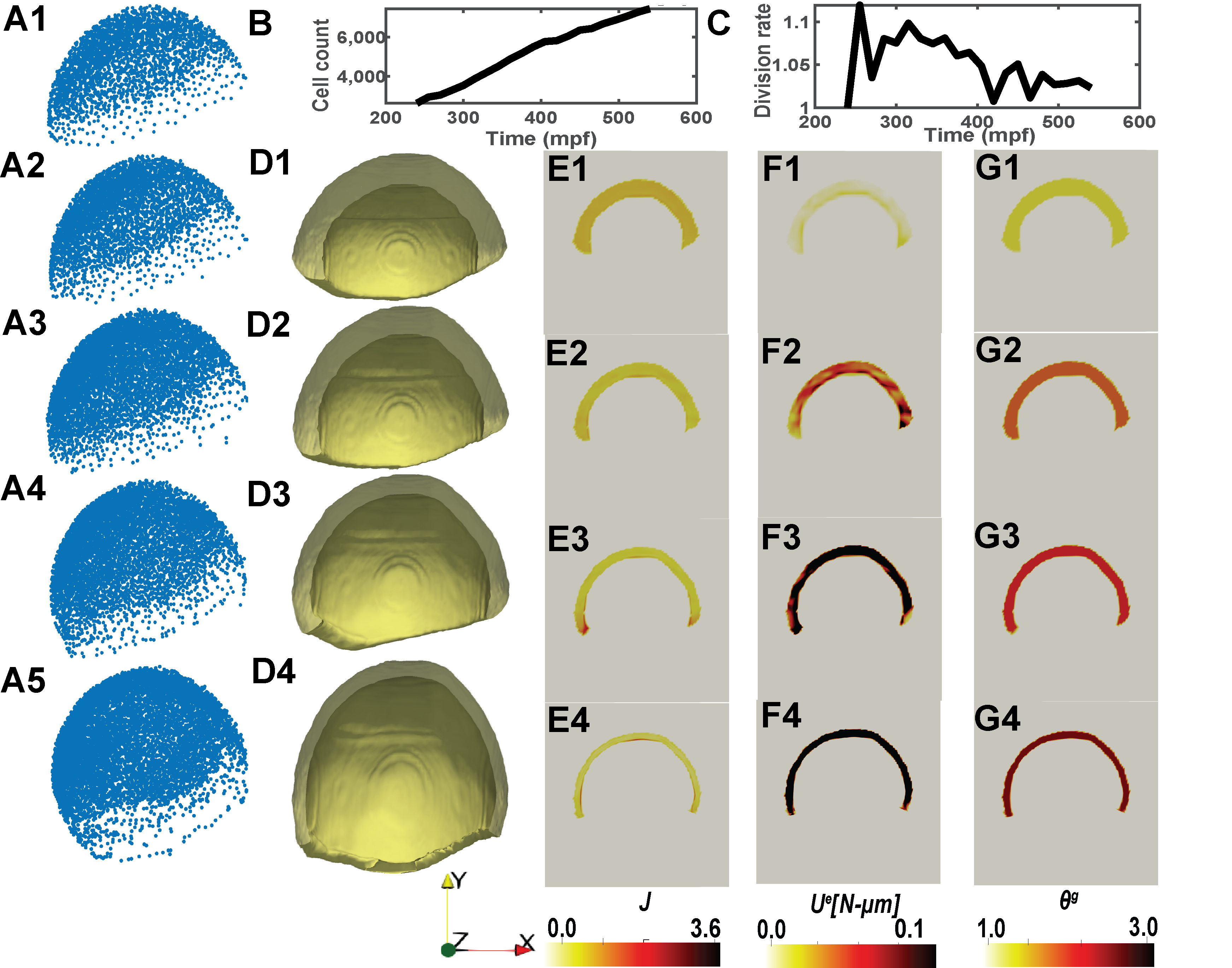}
\caption{Modeling of epiboly process. A1-A5:Quantitative cell position data obtain from whole mount live light sheet microscopy images of cell nuclei during epiboly \cite{keller2008reconstruction} at 241, 301, 361, 421 and 481 mpf. B: Nuclei number during epiboly from 241 mpf to 526 mpf. C: Cell division rate during epiboly. D1-D4: evolving implicit representation after registration shown at corresponding stages. E1-E4: Cross sections of the evolving meshes showing the strain energy contours from 241 mpf to 481 mpf predicted by the registration framework. F1-F4: Cross sections of the evolving meshes showing the local volume change as predicted by the registration framework. G1-G4: Cross sections of the evolving meshes showing the growth $\theta_g$ contours imposed based on the available data for cell division rates during epiboly.}
\label{fig:epiboly}
\end{figure} 

The cell position data in Fig. \ref{fig:epiboly}A1-A5 was first converted to a volumetric mesh and also an implicit representation for surface registration at different stages of the epiboly as shown in Fig. \ref{fig:epiboly}D1-D4. The resolution for the implicit representations of the geometries were $100\times100\times100$ pixels each. Not all time points in the data set were used. Surfaces were reconstructed for: 241, 301, 361, 421 and 481 mpf. The source geometry is the image obtained at 241 mpf and the target geometries are at the remaining time frames. The number of cell nuclei and cell division rate was calculated at each time frame as shown in Fig. \ref{fig:epiboly}B-C. For registration, the B-spline grid was initially set to $16\times16\times16$ elements and, during registration, three refinement levels were done. The Lame's parameters $\mu$ is set as $\SI{0.1}{\pascal}$ and $\lambda$ set as $\SI{1}{\pascal}$. $\alpha$ is set as $2$ and doubled every $2,000$ iterations. $\beta$ and $\beta_1$ are set as $1$ and $2.5$, respectively.

The total registration process is carried out in multiple stages, where the source geometry is fixed and the target geometry is updated with the corresponding time frames. In between the different stages we assume perfect plastic deformation and reset the control points while keeping the intermediate, registered image. The plastic deformation is assumed based on the growth of the embryo, which is imposed through the known increase in cell number at different time points during epiboly (see Fig. \ref{fig:epiboly}B). The volumetric change due to growth, $theta^g$, is assumed to be uniform throughout the volume of the tissue. We calculate $\theta^g_{n}$ at time $n$ as $\theta^g_n = {N_{cell}^{(n)}}/{N_{cell}^0}$, where $N_{cell}^{(n)}$ and $N_{cell}^(0)$ are the cell counts at time $n$ and 241 mpf, respectively.               

As seen from the registered meshes in Fig. \ref{fig:epiboly}D1-D4, we can observe the spreading and thinning of the embryonic tissue as the epiboly progresses. From the registered meshes, we can show that the proposed registration framework, by imposing the linear momentum residual as a driving force, predicts a spatially heterogeneous distribution of strains. The main results of the registration with this constraint are the contours for the strain energy ($U^{e}$) (Fig. \ref{fig:epiboly}E1-E4) and total volume change ($J$) (Fig. \ref{fig:epiboly}F1-F4) fields. We also plot the contours of growth ($\theta^g$) for the corresponding registered geometries at different stages of the epiboly in Fig. \ref{fig:epiboly}G1-G4. However, as mentioned, the growth field is not an output in this case, but rather it is imposed from the available data. From the local volumetric change contours, we can see that maximum deformation occurs near the the leading edge of epiboly. Remarkably, this is the region over which actin polymerization and myosin-driven contraction have been observed in experiments \cite{campinho2013tension, behrndt2012forces}. Therefore, our registration suggests that physically realistic deformations that match the overall shape changes in epiboly are those with increasing strain near the leading edge. Intuitively, past the equator point, increases in the area due to cell division cannot continue to spread over the yolk cell without some elastic deformation present. Mechanistically, the elastic strains needed to continue the spread of the EVL past the equator could come in part from the actomyosin ring at the leading edge of the EVL. These observations should of course be further refined and compare against new experimental data, but it should be highlighted how coupling the observed changes in geometry with a registration framework including some of the physics of the process offers a new tool to gain deeper insights into the fundamental mechanisms that control early embryo development.
\section{Tissue Expansion for Reconstructive Surgery}
\label{sec5}
Tissue expansion is a clinical procedure used to grow new skin \textit{in situ} which can be used in reconstructive surgeries \cite{logiudice2004pediatric}. Skin responds to sustained stretches by permanently increasing its area. The sustained stretches of skin are achieved by balloon-like devices called tissue expanders, which are inserted subcutaneously and inflated gradually over months \cite{radovan1984tissue}. Applications of this technique include breast reconstruction after mastectomy \cite{bertozzi2017tissue}, repair of large congenital defects  \cite{rivera2005tissue}, and skin grafting in burn patients \cite{chun1998versatility}. We have previously reported the development of an experimental model of tissue expansion in the swine \cite{lee2018improving}. Briefly, Yucatan minipigs were tatooed with four $10 \times 10\SI{}{cm^2}$ grids on the back. Tissue expanders of dimensions $4 \times 6\SI{}{cm^2}$ were placed in subcutaneous pockets while contralateral sides served as controls. In our previous work \cite{han2022bayesian}, we have fitted B-spline surfaces to the skin patches at different time points during the protocol, and used these B-spline surfaces to calculate the relative deformation during tissue expansion. However, to then calibrate our mechanobiological model of skin growth, we have separately built finite element models to solve the forward problem, i.e. how the tissue is deformed and grows in response to tissue expansion, see \cite{han2022bayesian}. The main limitation of the approach in \cite{han2022bayesian} is that the forward, finite element model, uses a simplified mesh of the skin and assumes it is a flat piece of tissue. Instead, the ideal analysis would be to simultaneously perform the registration and solving the PDEs for tissue deformation and growth. The registration framework presented here is the perfect tool for this application.    

\begin{figure}[!ht]
\centering
\includegraphics[width=1.01\textwidth]{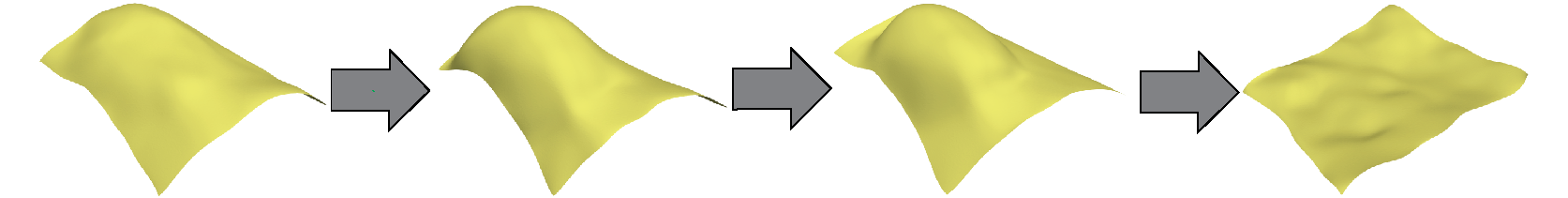}
\caption{Tissue expansion protocol. B-spline surfaces obtained at four configurations: before the dilation of the expander (\textit{pre-fill}), after dilation of expander (\textit{post-fill}), at the end of the tissue expansion protocol (\textit{TE-end}) and after skin excision (\textit{ex-vivo}).}
\label{fig:pigexpansion_protocol}
\end{figure}

The main stages of the protocol for the tissue expansion in the swine are shown in Fig. \ref{fig:pigexpansion_protocol}. The expander was dilated at 30 cc pressure in a single step over a period of 7 days. The stages of skin expansion were captured using three dimensional (3D) photographs. The 3D photographs were used to fit cubic B-spline surfaces \cite{han2022bayesian}. 
We start with the \textit{pre-fill} geometry which is the reference configuration. Note that this configuration is assumed to be a stress-free state since it is imaged at the start of the tissue expansion. The deformation of the \textit{pre-fill} stage to the \textit{post-fill} stage is assumed to be purely an elastic deformation, i.e. no growth,  because the post-fill image is taken immediately after inflation when the tissue has had no time to grow. Since the skin is a thin surface, in this application we are more interested in the area change after expansion. Given the unit surface normal $\mathbf{n}_{0}$ in the source geometry and the total deformation gradient $\mathbf{F}$, we evaluate the area change as

\begin{equation}
    \vartheta ={\lVert}\mathrm{cof}(\mathbf{F}).\mathbf{n}_{0}{\rVert}=\vartheta^{e}\, \vartheta^{g},
\label{eq_area_change}
\end{equation} where $\vartheta^{e}$ and $\vartheta^{g}$ are the elastic area change and growth area change, respectively. For the second stage, \textit{post-fill} to \textit{TE-end}, both elastic deformation and growth are considered. Thus, we introduce the growth tensor for area growth $\mathbf{F}_g = \sqrt(\vartheta^{g})\mathbf{I} + [1-\sqrt(\vartheta^{g})]\mathbf{n}_0 \otimes\mathbf{n}_0$ \cite{tepole2011growing,zollner2012biomechanics}. Update of the scalar growth variable is done with the same ODE as in Eqn. (\ref{eq_thetag_2}). 


Given the spline surfaces from \cite{han2022bayesian}, we add a thin layer of thickness $\SI{0.2}{\centi \meter}$ and construct an image as an implicit representation of the shape. We assume that the normal $\mathbf{n}_0$ is constant along the thickness and consistent with the normal map of the original B-spline surface. To ensure this, we invert the surface normal vectors at the bottom layer to keep them in the same direction as the top layer. Then, using Gaussian filter with standard deviation $\sigma = 1.5$, we interpolate the normal vector so that it is uniform across the thickness of the implicit representation.

\begin{figure}[!ht]
\centering
\includegraphics[width=\textwidth]{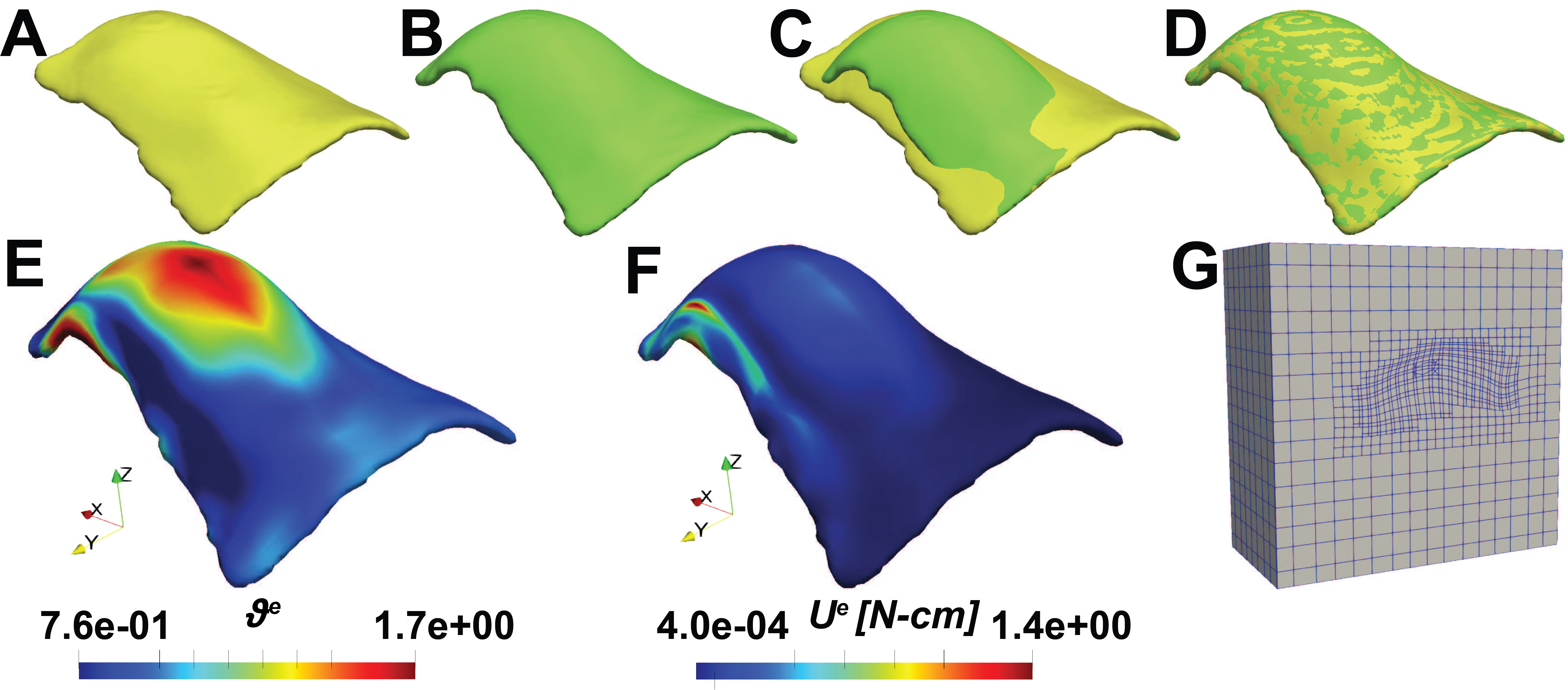}
\caption{Shape registration for the modeling of skin growth during tissue expansion between the \textit{pre-fill} and \textit{post-fill} stages. The contours of the implicit representation of the source  geometry (A), target geometry (B), initial overlap between the source and target geometry (C) and the overlap between the registered implicit function and the target geometry (D). The area change due to elastic deformation ($\vartheta^{e}$) which is also the area change due to total deformation is plotted over the implicit representation and is shown in E. No growth is considered here. The strain energy ($U^e$) is plotted over the implicit representation and is shown in F. THB-spline grid after 3 refinement levels is shown in G.}
\label{fig:prefill_postfill}
\end{figure} 

\begin{figure}[!ht]
\centering
\includegraphics[width=\textwidth]{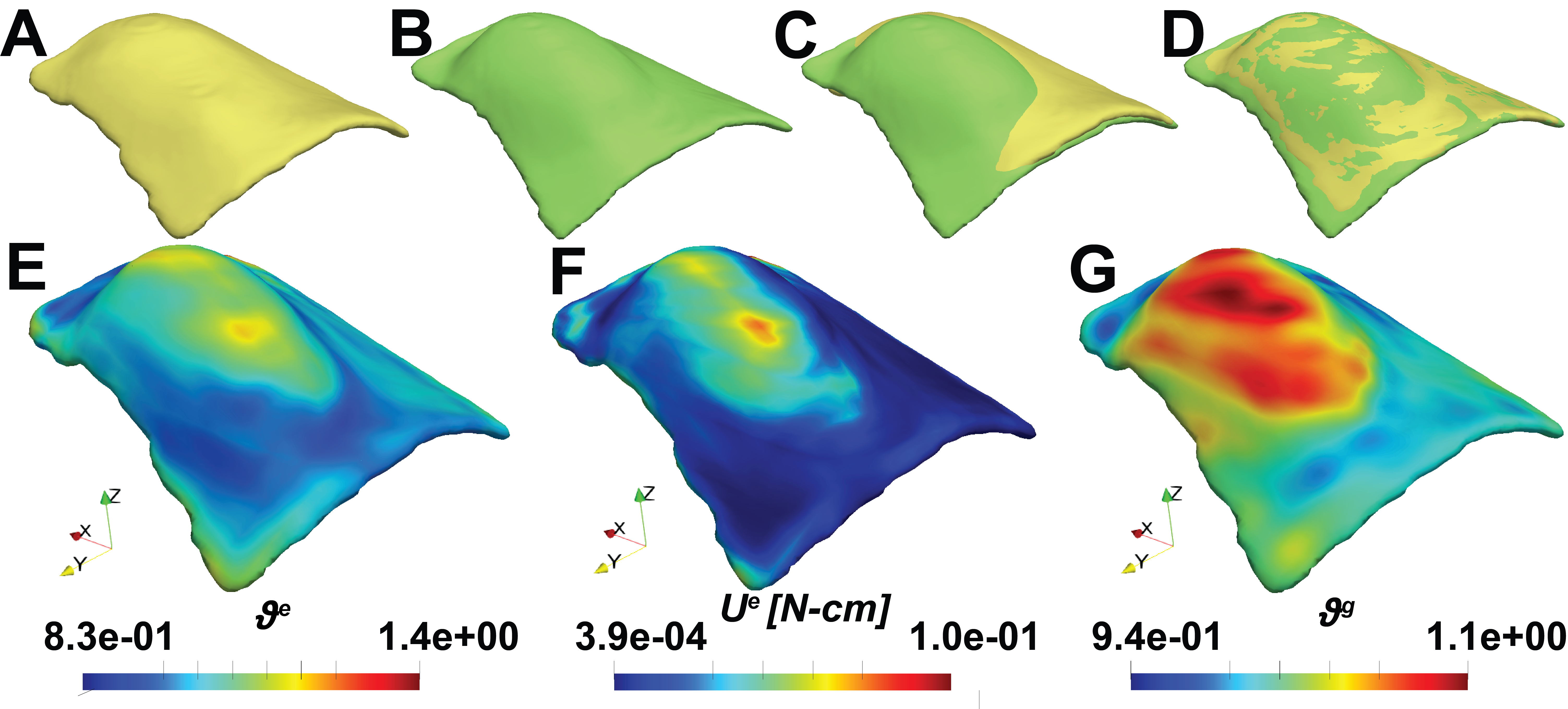}
\caption{Shape registration for the modeling of skin growth during tissue expansion between the \textit{pre-fill} and \textit{TE-end} stages. The contours of the implicit representation of the source  geometry (A), target geometry (B), initial overlap between the source and target geometry (C) and the overlap between the registered implicit function and the target geometry (D). The area change due to elastic deformation ($\vartheta^{e}$), the strain energy ($U^e$) and the area change due to growth ($\vartheta^g$) are shown in E, F and G, respectively.}
\label{fig:prefill_presac}
\end{figure}

In Fig. \ref{fig:prefill_postfill}, we perform surface registration from the \textit{pre-fill} to \textit{post-fill} stage. For this stage the grid resolution for the implicit function is $100\times100\times100$ pixels and the initial THB-spline control grid for registration has $16\times16\times16$ elements which are then locally refined up to 3 refinement levels during registration. $\alpha$ is set as $8$ and doubled every $375$ iterations whereas $\beta$ and $\beta_1$ are set as $1$ and $10$, respectively. Here since we only pure elastic deformation, we show the contours of the total area change ($\vartheta$) and the strain energy $U^{e}$. From the overlap of the registered geometry with the target geometry before and after registration in Fig. \ref{fig:prefill_postfill}C-D, we can see that the registration framework can accurately capture the target geometry even if there is a large and complex tissue deformation. From Fig. \ref{fig:prefill_postfill}E, we can see higher area change is observed near the apex of the expander. From Fig. \ref{fig:prefill_postfill}F, we also see that strain energy is higher not only close to the apex of the expander but near the skin patch boundary, which is also undergoing large deformation. The spatial distribution of the area change over the skin matches our previous work \cite{lee2018improving,han2022bayesian}.



In Fig. \ref{fig:prefill_presac}, we demonstrate the results from the surface registration framework  from the \textit{pre-fill} stage all the way to the \textit{TE-end} stage. Registration is carried out in two steps. We first deform the source image of the \textit{pre-fill} to a first target, the \textit{post-fill} image, considering only elastic deformation as shown before. But then we continue with the registration and change the target to the \textit{TE-end} geometry, and we consider both elastic and growth deformations in this second stage. We evaluate the registration for two refinement levels at the second stage with an initial grid of $12\times12\times12$ elements. The growth rate $\kappa$ is set as $\SI{0.018}{\hour^{-1}}$ and $\vartheta^{crit}$ is set as $1$. Since it is difficult to evaluate interpolated stages between the source and target geometry for this example, we set $\alpha$ as $0.1$ initially and slowly increase it by $0.05$ every 10 iterations. Growth update is also carried out every 10 iterations over 7 days. We plot the elastic and growth area change along with the strain energy contours. Here, we can see that both the elastic and growth area changes are higher near the apex of the expander. This is expected as the maximum stretch occurs near the apex, thus, integration of Eqn. (\ref{eq_thetag_2}) results in more skin growth in this region. This is again consistent with our previous work \cite{han2022bayesian}. 


\section{Discussion and Conclusions}
\label{sec_conclusion}
In this article, we propose a novel shape registration framework to capture the deformation of biological tissues from imaging data while satisfying linear momentum balance and accounting for permanent deformations due to growth. The problem is set up as a strain energy minimization problem with a penalty for the image mismatch. The deformation of the image is done using THB-splines with local refinement. The control points of the THB-spline grid are updated in a dynamic relaxation scheme based on a variation of the energy. The variational approach effectively produces a residual vector for internal forces due to the tissue deformation, and a residual vector of external forces from the image mismatch. After checking that the algorithm worked as intended on benchmark examples, we applied the shape registration framework to study the growth of biological tissues in two applications: tissue expansion and zebrafish embryo epiboly. 

Registration frameworks that can account for physical phenomena are being actively developed for a range of applications such as tumor growth and mapping of cardiac strains in the beating heart \cite{bauer2010atlas,genet2018equilibrated}, to name a couple of examples. Here, our contribution is on combining the registration problem with the consideration of linear momentum balance and tissue growth. The theoretical framework for tissue growth used here follows a long stride of developments since the introduction of the multiplicative split of the deformation gradient to model growth by Rodriguez et al. \cite{rodriguez1994stress}. Over the past couple of decades, this multiplicative split into growth and elastic deformations has been widely used to model the growth of a large class of tissues, e.g. airways \cite{eskandari2015patient}, heart \cite{goktepe2010generic}, and the brain \cite{budday2014mechanical}. Numerically, modeling of growing soft tissues has been done primarily with the finite element method \cite{eskandari2015systems}. We show the corresponding implementation in the context of isogeometric analysis frameworks, which have gained increasing popularity due to the high continuity of basis functions \cite{giannelli2012thb}. The multiplicative growth framework is certainly not the only theory to describe the evolving mechanics of living matter. For instance, mixture theory approaches are an alternative formulation that could be incorporated with our framework in the future \cite{valentin2013finite}. 

Another key feature of the proposed framework is the use of THB-splines with local refinement. We have done extensive work on registration with THB-splines \cite{pawar2016adaptive}, where we have shown that the isogeometric framework allows for smooth registration mappings and can handle extreme deformations. The use of THB-splines allows for local refinement and an efficient numerical implementation. These features are maintained in the formulation shown here. The code is made publicly available through the Github link at the end of the article. 

There are certainly some limitations of the framework. The dynamic update is such that it eventually converges to satisfy linear momentum balance and approaches the desired registered image. However, because the dynamic relaxation arises from gradient-descent of competing energy terms instead of the imposition of hard constraints, there is no exact satisfaction of the image alignment problem. In other words, the image mismatch is penalized, but cannot be driven to be exactly zero. This limitation of dynamic relaxation methods for image registration is common beyond the method shown here \cite{leng2013medical,jia2015novel,pawar2018dthb3d,pawar2019joint}. Nonetheless, as the image mismatch penalty is increased over several iterations, the source image is adequately mapped onto the target as shown in our examples.  Another limitation of the dynamic relaxation approach is that it is sensitive to the parameters that control the relative weight between the image and hyperelastic residuals. As shown, when the image residual is scaled to be large relative to the hyperelastic residual, the deformation of the grid is physically unrealistic. Even though, even in such cases the continuous relaxation of the elastic energy should eventually converge to the equilibrium solution, this can take a large number of iterations. Furthermore, since an explicit scheme is used, the time step is subjected to stability considerations. We explored the relative scaling of the image and hyperelastic residuals to inform our choice of parameters our examples, but a change in the formulation, away from the dynamic relaxation approach and towards the solution of the equilibrium problem would be one alternative to deal with existing limitation. Nevertheless, by manually tuning the parameters, we were able to get excellent registration results within a few hundred iterations. 

The real-world impact of the framework is showcased in the two application examples. Zebrafish embryo development is an ideal biological system to understand the mechanical cues that lead to morphogenesis. Given growth measured from cell division and the images of the evolving embryo shape from light-sheet microscopy, we used our registration framework to predict elastic deformation profiles during epiboly. Interestingly, our prediction is that the elastic deformation is not homogeneous. Instead, the analysis predicts that elastic deformations are needed at the leading edge of the EVL as it passes the equator and continues to engulf the yolk cell. This observation aligns with the existence of an actomyosin ring which is thought to contract at the leading edge of the EVL guiding the overall shape change of the embryo \cite{campinho2013tension,behrndt2012forces}. The other application shown here is skin growth in tissue expansion. We have done work characterizing skin growth on a porcine animal model \cite{lee2018improving}. For the porcine model we tattoo grids of the backs of the animals, which allows us to easily reconstruct B-spline surfaces \cite{tepole2017quantification}. However, to being able to translate the analysis to the clinical setting it is indispensable to have a registration framework that works in the absence of tattooed grids. Additionally, in our previous work we have calibrated our models of skin growth using finite element models of idealized geometries \cite{han2022bayesian}. The registration method in the present allows us to do the analysis of skin growth in the same geometries that are available from 3D photography instead of the simplified models. We found more deformation at the apex, which have independently seen in our previous work, and we consequently predict more growth at apex, which is also consistent with our finite element model of the idealized geometry \cite{lee2018improving,han2022bayesian}. 

In conclusion, we anticipate that our PDE-constrained shape registration method accounting for growth of living tissue will offer a new tool to the community to better understand the adaptation of tissues to mechanical cues in development, health and disease.
\backmatter

\bmhead{Data Availability}
Associated files to this publication are available at \url{https://github.com/arpawar/StrainMin_Registration}

\bmhead{Acknowledgments}
The research at Purdue University was supported in part by the NIHR01GM132501-02 and NIHR01AR074525-01A1.


\end{document}